\documentclass[runningheads]{llncs}
\usepackage{graphicx}
\usepackage{amsmath}
\usepackage{multirow}
\usepackage{array}
\usepackage{adjustbox}
\usepackage{xcolor}
\newcommand{\rpm}{\raisebox{.2ex}{$\scriptstyle\pm$}}
\begin{document}
\title{A Matrix Autoencoder Framework to Align the Functional and Structural Connectivity Manifolds as Guided by Behavioral Phenotypes}
%
%

\authorrunning{Niharika Shimona D'Souza et al.} 
%
%
\author{Niharika Shimona D'Souza\inst{1} 
 \textsuperscript{*}
\and Mary Beth Nebel~\inst{2}~\inst{3} \and Deana Crocetti~\inst{2} \and Joshua Robinson \inst{2} \and Stewart Mostofsky\inst{2}\inst{3}\inst{4} \and Archana~Venkataraman\inst{1} }


\institute{Dept. of Electrical and Computer Eng., Johns Hopkins University, Baltimore, USA
\email{\textsuperscript{*}}{Shimona.Niharika.Dsouza@jhu.edu}
\and
Center for Neurodevelopmental \& Imaging Research, Kennedy Krieger Institute
\and
Dept. of Neurology, Johns Hopkins School of Medicine, Baltimore, USA
\and
Dept. of Psychiatry \& Behavioral Science, Johns Hopkins School of Medicine, USA}
\maketitle              
\begin{abstract}
We propose a novel matrix autoencoder to map functional connectomes from resting state fMRI (rs-fMRI) to structural connectomes from Diffusion Tensor Imaging (DTI), as guided by subject-level phenotypic measures. Our specialized autoencoder infers a low dimensional manifold embedding for the rs-fMRI correlation matrices that mimics a canonical outer-product decomposition. The embedding is simultaneously used to reconstruct DTI tractography matrices via a second manifold alignment decoder and to predict inter-subject phenotypic variability via an artificial neural network. We validate our framework on a dataset of 275 healthy individuals from the Human Connectome Project database and on a second clinical dataset consisting of 57 subjects with Autism Spectrum Disorder. We demonstrate that the model reliably recovers structural connectivity patterns across individuals, while robustly extracting predictive and interpretable brain biomarkers in a cross-validated setting. Finally, our framework outperforms several baselines at predicting behavioral phenotypes in both real-world datasets.
\end{abstract}

\section{Introduction}
The brain is increasingly viewed as an interconnected network. Two key elements of this network are the structural pathways between brain regions and the functional signaling that rides on top. This structural connectivity information can be measured via Diffusion Tensor Imaging (DTI) tractography~\cite{assaf2008diffusion,le2001diffusion}. Likewise, resting-state fMRI (rs-fMRI) captures inter-regional co-activation, which can be used to infer functional connectivity~\cite{lee2013resting,fox2007spontaneous}. Several studies have found both direct and indirect correspondences between structural and functional connectivity~\cite{honey2009predicting,fukushima2018structure}. Going a step further, structural and functional connectivity have been shown to be predictive of each other at varying scales \cite{messe2015predicting,chu2018function,zhang2020recovering}. Hence, multimodal integration of these viewpoints has become a key area of focus for characterizing neuropsychiatric disorders such as autism and ADHD~\cite{liu2015multimodal,zhang2020deep}.

\par In the clinical neuroscience realm, techniques for integrating structural and functional connectivity focus on group-wise discrimination. These works include statistical tests on edge-based features to identify significant differences in Alzheimer's disease~\cite{hahn2013selectively}, parallel ICA using structure and function to identify discriminative biomarkers of schizophrenia~\cite{sui2013combination}, and classical machine learning techniques to predict diagnosis~\cite{castellazzi2020machine}. While highly informative at a group level, these methods do not directly address inter-individual variability, for example by predicting finer grained patient characteristics. This divide has been partially bridged by end-to-end deep learning models. Examples include MLPs \cite{lin2016predicting} for age prediction from functional connectomes and convolutional neural networks \cite{kawahara2017brainnetcnn} for predicting cognitive and motor measures from structural connectomes. The work of \cite{d2019integrating} takes the alternate approach of combining a dictionary learning model on the functional connectomes coupled with an Artificial Neural Network (ANN) that predicts multiple clinical measures, while also preserving interpretability. Even so, these models focus exclusively on a single neuroimaging modality and do not exploit the interplay between function and structure.

\par Geometric learning frameworks have recently shown great promise in multimodal connectomics studies, both for conventional manifold learning~\cite{wong2018riemannian} and in the context of Graph Convolutional Networks (GCN) \cite{liu2019community,zhang2020deep}. Their primary advantage is the ability to directly incorporate and exploit the underlying data geometry. Beyond associative analyses, the work of \cite{zhang2020recovering,bessadok2020topology} employ multi-GCNs combined with a Generative Adversarial Network (GAN) for the alignment problem. Particularly, \cite{zhang2020recovering} examines the problem of recovering structural connectomes from patient functional connectomes  While this paper marks a seminal contribution to multimodal integration, the representations learned by end-to-end GCNs can be hard to interpret. It can also be difficult to train GANs on modest-sized datasets \cite{karras2020training}.
\par We propose an end-to-end matrix autoencoder  that maps rs-fMRI correlation matrices to structural connectomes obtained from DTI tractography. Inspired by recent work in Riemannian deep learning \cite{huang2017riemannian,dong2017deep}, our matrix autoencoder, estimates a low dimensional embedding from rs-fMRI correlation matrices while taking into account the geometry of the functional connectivity (FC) manifold. Our second matrix decoder uses this embedding to reconstruct patient structural connectivity (SC) matrices akin to a manifold alignment \cite{wang2008manifold} between the FC and SC data spaces. For regularization, the FC embedding is also used to predict behavioral phenotypes. We demonstrate that our framework reliably traverses from function to structure and extracts meaningful brain biomarkers.
\section{A Matrix Auto-Encoder for Connectome Manifolds}
\begin{figure}[t!]
   \centering
   \includegraphics[width=\dimexpr \textwidth-14\fboxsep-10\fboxrule\relax]{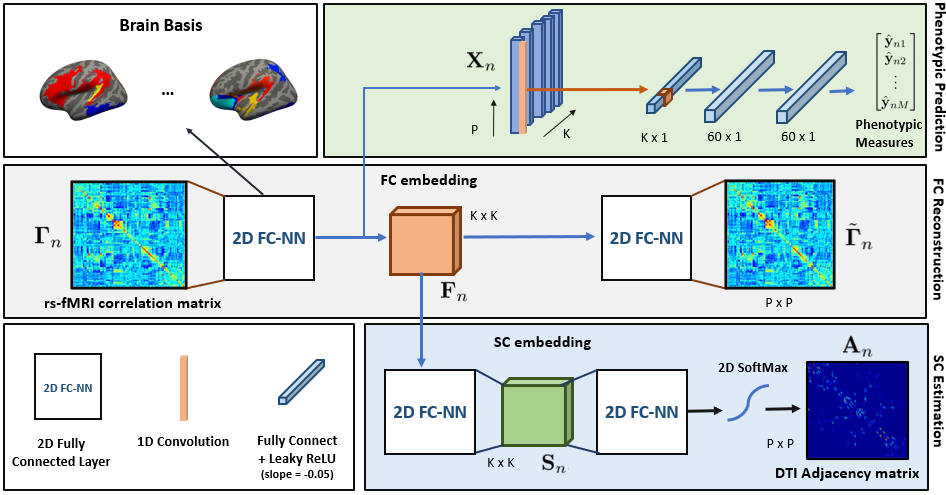}
   \caption{A Matrix Autoencoder for aligning the FC and SC manifolds \textbf{Gray Box:}~Matrix encoder-decoder for functional connectomes. \textbf{Blue Box:}~Alignment Decoder for estimating DTI connectomes \textbf{Green Box:}~ANN for predicting behavioral phenotypes}  
   \label{MAE}
\end{figure}
Fig.~\ref{MAE} illustrates our matrix autoencoder framework consisting of an encoder-decoder for functional connectivity (gray box), manifold alignment for estimating structural connectivity (blue box), and ANN for prediction of behavioral phenotypes (green box). Let $N$ be the number of patients and $P$ be the number of ROIs in our brain parcellation. We denote the rs-fMRI correlation matrix for patient $n$ by $\mathbf{\Gamma}_{n}\in \mathcal{R}^{P \times P}$. $\mathbf{A}_{n} \in \mathcal{R}^{P \times P}$ is the corresponding structural connectivity profile, and $\mathbf{y}_{n} \in \mathcal{R}^{M \times 1}$ is a vector of $M$ concatenated phenotypic measures.

\medskip
\noindent{\underline{\textbf{Functional Connectivity Reconstruction:}}} By construction, the correlation matrices $\mathbf{\Gamma}_{n}$ belong to the manifold of symmetric positive semi-definite matrices $\mathcal{P}^{+}_{P}$. Our matrix autoencoder estimates a latent functional embedding $\mathbf{F}_{n} \in \mathcal{P}_{K}^{+}$ using a 2D fully connected (2D FC-NN) layer~\cite{dong2017deep,huang2017riemannian}. Formally, this mapping $\mathbf{\Phi}_{\text{ec}}(\cdot): \mathcal{P}^{+}_{P} \rightarrow \mathcal{P}^{+}_{K}$ is parametrized by weights $\mathbf{W} \in \mathcal{R}^{P \times K}$ and is computed as a cascade of two linear layers with tied weights: $\mathbf{F}_{n} = \mathbf{\Phi}_{\text{ec}}(\mathbf{\Gamma}_{n}) = \mathbf{W}^{T}\mathbf{\Gamma}_{n}\mathbf{W}$. Our decoder is another 2D FC-NN that estimates $\tilde{\mathbf{\Gamma}}_{n}$ from $\mathbf{F}_{n}$ via a similar transformation $\mathbf{\Phi}_{\text{dc}}(\cdot): \mathcal{P}^{+}_{K} \rightarrow \mathcal{P}^{+}_{P}$ that shares weights with the encoder. Mathematically, our FC reconstruction loss is represented as follows:
\noindent
\begin{align}
\mathcal{L}_{\text{FC}} = \frac{1}{N}\sum_{n}{\vert\vert{\mathbf{\Phi}_{\text{dc}}(\mathbf{\Phi}_{\text{ec}}(\mathbf{\Gamma}_{n}))- \mathbf{\Gamma}_{n}}\vert\vert}^{2}_{F} = \frac{1}{N}\sum_{n}{\vert\vert{\mathbf{W}\mathbf{W}^{T}\mathbf{\Gamma}_{n}\mathbf{W}\mathbf{W}^{T}- \mathbf{\Gamma}_{n}}\vert\vert}^{2}_{F}
\label{eqn:FC}
\end{align}
The second term of Eq.~(\ref{eqn:FC}) encourages the columns of the brain basis $\mathbf{W}$ to be orthonormal. Conceptually, this specialized loss helps us learn uncorrelated patterns that explain the rs-fMRI data well while acting as an implicit regularizer.

\medskip
\noindent{\underline{\textbf{Structural Connectivity Estimation:}}} The structural connectivity matrices $\mathbf{A}_{n}$ are derived from DTI tractography and belong to the manifold of symmetric (non PSD) matrices $\mathcal{S}_{P}$. Our alignment decoder first generates an SC embedding $\mathbf{S}_{n} \in \mathcal{R}^{K\times K}$ from $\mathbf{F}_{n}$ via a 2D FC-NN layer $\mathbf{\Phi}_{\text{align}}(\cdot):\mathcal{P}^{+}_{K} \rightarrow \mathcal{P}^{+}_{K}$, followed by a second 2D FC-NN layer $\mathbf{\Phi}_{\text{est}}(\cdot):\mathcal{P}^{+}_{K} \rightarrow \mathcal{P}^{+}_{P}$ which maps to the structural connectivity matrices. For stability our SC matrices do not have self-connections and  are normalized to $\lVert \mathbf{A}_{n} \rVert_1 = 1$. Accordingly, at the output layer, we apply a 2D softmax $\mathcal{SF}(\cdot)$,  and then suppress the diagonal elements to generate the final output $\tilde{\mathbf{A}}_{n} \in \mathcal{R}^{P \times P}$. Our SC estimation objective is represented as follows:

\noindent
\begin{align}
\mathcal{L}_{\text{SC}} = \frac{1}{N}\sum_{n}{\Big\vert\Big\vert{\mathcal{SF}\Big[\mathbf{\Phi}_{\text{est}}(\mathbf{\Phi}_{\text{align}}(\mathbf{F}_{n})) \circ [\mathbf{1}\mathbf{1}^{T} -\mathcal{I}_{P}]\Big] - \mathbf{A}_{n}}\Big\vert\Big\vert}^{2}_{F} 
\label{eqn:SC}
\end{align}
where $\circ$ is the element-wise Hadamard product. $\mathbf{1}\in\mathcal{R}^{P\times 1}$ is the vector of all ones, and $\mathcal{I}_{P}$ is the identity matrix of dimension $P$. Conceptually, the loss in Eq.~(\ref{eqn:SC}) is akin to manifold alignment \cite{wang2008manifold} between the functional and structural embeddings based on a two sided Procrustes-like objective.

\medskip
\noindent{\underline{\textbf{Phenotypic Prediction:}}}
We map the intermediate representation $\mathbf{X}_{n}=\mathbf{\Gamma}_{n}\mathbf{W}$ $\in \mathcal{R}^{P \times K}$ learned by the FC encoder to the phenotypes $\mathbf{y}_{n}$ via a cascade of a 1D convolutional layer and an ANN. The convolutional layer $\mathcal{F}_{\text{conv}}(\cdot)$ collapses $\mathbf{X}_{n}$ along its rows via a weighted sum to generate a $K$ dimensional feature vector. This feature vector is input to a simple two layered ANN $\mathcal{G}(\cdot)$ to jointly estimate the elements in $\hat{\mathbf{y}}_{n}$. We use a Mean Squared Error (MSE) loss function:
\noindent
\begin{align}
    \mathcal{L}_{\text{phen}} = \frac{1}{NM} \sum_{n}{\vert\vert{\hat{\mathbf{y}}_{n} - {\mathbf{y}}_{n}}\vert\vert}^{2}_{F} = \frac{1}{NM} \sum_{n} {\vert\vert{\mathcal{G}({\mathcal{F}_{\text{conv}}(\mathbf{X}_{n}))- \mathbf{y}_{n}}}\vert\vert}^{2}_{2}
    \label{eqn:behav}
\end{align}
This prediction task is a secondary regularizer that encourages our matrix autoencoder to learn representations predictive of inter-subject variability.

\medskip
\noindent{\underline{\textbf{Implementation Details:}}} We train our framework on a  joint objective that combines Eqs.~(\ref{eqn:FC}), (\ref{eqn:SC}) and (\ref{eqn:behav}) as follows:
\begin{align}
\mathcal{L} = \mathcal{L}_{\text{FC}} + \gamma_{1} \mathcal{L}_{\text{SC}} + \gamma_{2} \mathcal{L}_{\text{phen}}
\label{eqn:comb}
\end{align}
where $\gamma_{1}$ and $\gamma_{2}$ balance the tradeoff for the SC estimation and phenotypic prediction relative to the FC reconstruction objective. 
We employ an ADAM optimizer \cite{kingma2014adam} with learning rate $0.005$ and weight decay regularization \cite{loshchilov2017decoupled} ($\delta = 0.0005$) run for a maximum of $400$ epochs. Optimization parameters were fixed based on a validation set consisting of $30$ additional patients from the HCP database. We use this strategy to set the the dimensionality of our autoencoder embedding at $K=15$ and loss penalities to $\{\gamma_{1},\gamma_{2}\} = 10^{3},3$. Finally, we utilize a spectral initialization for the encoder-decoder weights $\mathbf{W}$ in Eq.~(\ref{eqn:FC}) based on the top $K$ eigenvectors of the average patient correlation matrix $\bar{\mathbf{\Gamma}}_{n}$ for the training set. We use a similar initialization based on $\bar{\mathbf{A}}_{n}$ for $\mathbf{\Phi}_{\text{est}}(\cdot)$, and default initialization \cite{lecun2012efficient} for the remaining layers. Our model has a runtime of 10-12 minutes on an $8$ core machine with $32$GB RAM implemented in PyTorch (v1.5.1). 

\subsection{Baseline Methods for Phenotypic Prediction}

\medskip \noindent
\textbf{Matrix AE without rs-fMRI decoder:} We start with the architecture in Fig.~\ref{MAE} but omit the rs-fMRI decoder loss ($\mathcal{L}_{\text{FC}}$) in Eq.~(\ref{eqn:comb}). This helps us evaluate the benefit of a tied encoder-decoder model for the rs-fMRI matrices.

\medskip \noindent
\textbf{Matrix AE without DTI decoder:} We start with the architecture in Fig.~\ref{MAE} but remove the DTI decoder loss ($\mathcal{L}_{\text{SC}}$) in Eq.~(\ref{eqn:comb}). This helps us evaluate the benefit of manifold alignment to constrain the functional embedding.

\medskip \noindent
\textbf{Decoupled Matrix AE + ANN:} We start with the architecture in Fig.~\ref{MAE} but decouple the representation learning on the connectomics data from the prediction of phenotypic measures by training the models separately. 

\medskip \noindent
\textbf{BrainNetCNN:} This baseline integrates multimodal connectivity data via the BrainNetCNN \cite{kawahara2017brainnetcnn}. We modify the original architecture, which is designed for a single modality, to have two branches, one for the rs-fMRI correlation matrices $\mathbf{\Gamma}_{n}$, and another for the DTI connectomes $\mathbf{A}_{n}$. The ANN is modified to output $M$ measures of clinical severity. We set the hyperparameters according to 
\cite{kawahara2017brainnetcnn}

\medskip \noindent
\textbf{rs-fMRI Dictionary Learning + ANN:} The framework in \cite{d2019integrating} uses rs-fMRI correlation matrices for the prediction of multiple clinical measures. The model combines a dictionary learning with a neural network predictor, with these two blocks optimized in an end-to-end fashion via a coupled optimization objective. 

\section{Experimental Evaluation and Results}

\medskip\noindent{\textbf{HCP Dataset and Preprocessing:}}
We download rs-fMRI and DTI scans for of $275$ healthy individuals from the Human Connectome Project (HCP) S1200 database \cite{van2013wu}. Rs-fMRI data was pre-processed according to the standard HCP pipeline \cite{smith2013resting}, which accounts for motion and physiological confounds. We selected a $15$-minute interval from the rs-fMRI scans to remain commensurate with standard rs-fMRI protocols. We used the Neurodata MR Graphs package \cite{kiar2016ndmg} to pre-process the DTI scans and estimate fiber bundles via streamline tractography. Our phenotypic measure was the Cognitive Fluid Intelligence Score (CFIS) \cite{duncan2005frontal,bilker2012development} adjusted for age, which is obtained via a battery of tests measuring cognitive reasoning (dynamic range:~$70-150$). We used the Automatic Anatomical Labelling (AAL) atlas with $116$ ROIs as our parcellation. From here, we compute the the Pearson correlation between the the mean rs-fMRI time course in the two regions; we subtract the contribution of the top (roughly constant) eigenvector to obtain the input $\mathbf{\Gamma}_n$. The input $\mathbf{A}_n$ are obtained by dividing the number of tracts between ROIs by the total fiber tracts across all ROI pairs.

\medskip\noindent{\textbf{Predicting Behavioral Phenotypes:}}
Table~\ref{table:HCP} (and Fig.~$1$ in Supplementary) compares the model against the baselines when predicting CFIS in a five-fold cross validated setting. Lower Median Absolute Error (MAE) and higher Normalized Mutual Information (NMI) and correlation coefficient (R) signify improved performance. Our framework outperforms the baselines during testing, though the model of \cite{d2019integrating} comes in a close second. This suggests that the Matrix Autoencoder faithfully models subject-specific variation even in unseen patients.

\begin{figure}[!b]
 \centerline{\includegraphics[scale=0.50]{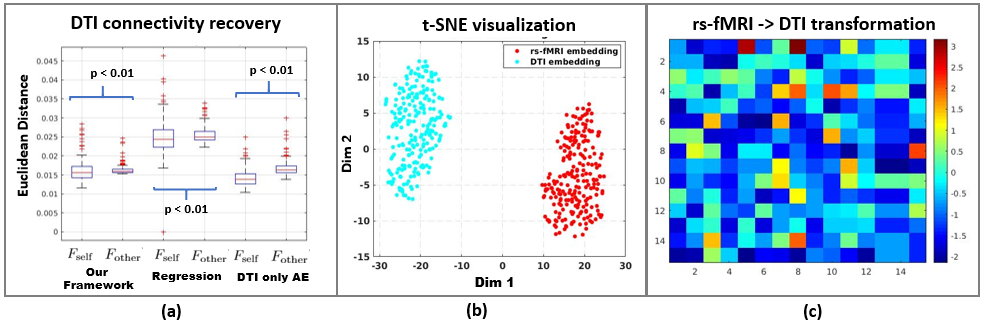}}
{\caption{\textbf{(a)} Recovery of SC for \textbf{(L):} Our Framework \textbf{(M):} Linear Regression \textbf{(R):} DTI only Autoencoder \textbf{(b)} t-SNE visualization for FC and SC embeddings \textbf{(c)} Coefficient of Variation $(C_{v})$ (log scale) for the weights of $\mathbf{\Phi}_{\text{align}}(\cdot)$. Cold colors imply small deviations, i.e. better stability}\label{Results}}
\end{figure}

\medskip\noindent{\textbf{Functional to Structural Association:}} We evaluate three aspects of our functional to structural manifold alignment. First is our ability to recover structural connectivity matrices during testing. Here, we compare two distance metrics: (1) $F_\text{self}$ is the Frobenius norm between a test example $\mathbf{A}_{n}$ and the model prediction for the same example $\hat{\mathbf{A}}_{n}$, and (2) $F_\text{other}$ is $\hat{\mathbf{A}}_{n}$ and other SC matrices $\mathbf{A}_{m}, (m \neq n)$. As shown in the left of Fig.~\ref{Results}(a), $F_\text{self}$ is consistently smaller than $F_\text{other}$, with statistical significance determined using the Wilcoxon rank sum test. This indicates that individual differences in SC are preserved by our framework. In the same plot, we also benchmark the recovery performance of our framework against a baseline Matrix encoder-decoder (gray box in Fig.~\ref{MAE}) with the SC matrices as \textit{input and output}. We also compare against a linear regression between the vectorized upper diagonal FC features (input) and SC features (output) to help evaluate the benefit of our matrix decomposition. As seen, our function $\rightarrow$ structure decoding achieves similar performance as directly encoding/decoding the structural connectivity. At the same time, the linear regression baseline performs worse than both of these techniques. This suggests that the ability to directly leverage the low rank matrix structure is key to preserving individual differences during reconstruction.
\begin{table*}[t!]
\centering
{{\caption{{CFIS prediction on the HCP dataset against the baselines using Median Absolute Error (MAE), Normalized Mutual Information (NMI) for training and testing, and Correlation Coefficient (R) for the test set. Best performance is highlighted in bold. \label{table:HCP}}}}{
\begin{tabular}{|c|c|c|c|c|c|} 
\hline 
  \textbf{Method} &\textbf{MAE Train } & \textbf{MAE Test} & \textbf{NMI Train} & \textbf{NMI Test} & \textbf{R} \\ 
\hline 
  No rs-fMRI dec. & 6.31~\rpm~{5.61} & 16.42~\rpm~{12.41} & 0.85 & 0.61 & 0.07  \\
  No DTI dec. & 6.30~\rpm~{5.80} & 15.44~\rpm~{13.00} & 0.86 & 0.61 & 0.11  \\
  Decoupled. & 2.53~\rpm~{2.41} & 14.90~\rpm~{13.60} & 0.87 & 0.59  & 0.10 \\
  BrainNetCNN & 6.80~\rpm~{6.25} & 14.95~\rpm~{12.74} & 0.88 & 0.59  & 0.12 \\
  Dict. Learn.~+~ANN & \textbf{3.19~\rpm~{2.19}} & 15.26~\rpm~{13.99} & \textbf{0.89} & 0.66 & 0.29  \\
  \textbf{Our Framework} & \underline{3.19~\rpm~{2.47}} & \textbf{14.08~\rpm~{11.85}} & 0.86 & \textbf{0.69} & \textbf{0.30}  \\
\hline
\end{tabular}
}}
\end{table*}

Second, we use t-SNE to visualize the symmetric FC and SC embeddings, $\mathbf{F}_{n}$ and $\mathbf{S}_{n}$, respectively. Fig.~\ref{Results}(b) displays the 2D t-SNE representation computed from the upper-triangle entries of the embedding. As seen, the FC and SC are clustered in two different locations within this space. Interestingly, the learned representations are non-overlapping without explicit enforcement. This suggests that the alignment decoder $\mathbf{\Phi}_{\text{align}}(\cdot)$ is learning a conversion between manifolds. 

Third, we examine the stability of the transformation learned by the alignment decoder, i.e. the weights $\mathbf{W}_{\text{align}} \in \mathcal{R}^{K \times K}$ of $\mathbf{\Phi}_{\text{align}}(\cdot)$. We first match the columns of $\mathbf{W}_{\text{align}}$ across cross validation folds according to correspondences between the functional brain basis. For each entry of $\mathbf{W}_{\text{align}}$, we compute the coefficient of variation ($C_{v}$), i.e. the ratio of the standard deviation to the mean (in absolute value). Lower values of $C_{v}$ indicate smaller deviations from the mean values, i.e. better stability. Fig.~\ref{Results}(c) displays the log coefficient of variation $\log (C_{v})$, where the cool colors indicate smaller $C_{v}$. As seen, a majority of the entries of $\mathbf{W}_{\text{align}}$ have low variation over the mean pattern value. Overall, our results suggest that our framework learns a stable mapping across the manifolds that explains individual patterns of structural connectivity faithfully.
\begin{figure}[t!]
 \centerline{\includegraphics[width=\dimexpr \textwidth-24\fboxsep-24\fboxrule\relax]{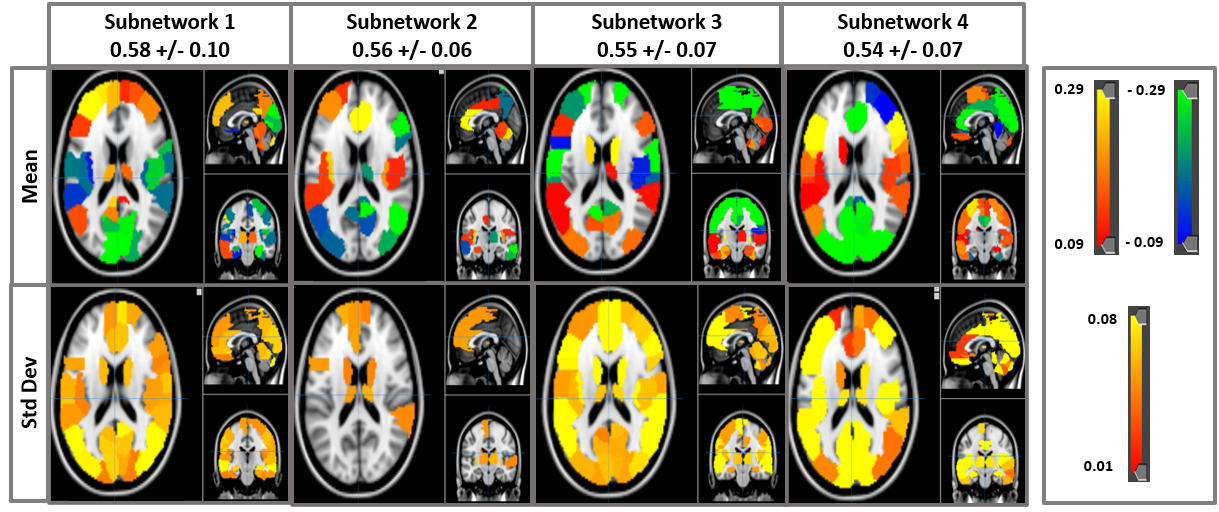}}
{\caption{Top four bases learned by the Matrix Autoencoder measured by the absolute correlation coefficient across cross validation folds and initializationa. }\label{Subnetworks}}
\end{figure}

\medskip\noindent{\textbf{Examining the FC Biomarkers:}} We explore the functional connectivity patterns learned by our framework by first matching the brain bases (i.e., columns of $\mathbf{W}$) across the cross validation folds based on the absolute correlation coefficient. We run this experiment five times with different initializations for the ANN branch to check for consistency in the learned representation. Fig~\ref{Subnetworks} displays the four most consistent bases, as projected onto the brain using the region definitions of the AAL atlas.
We notice that while there is spatial overlap between the bases, the standard deviations are small, which indicates that our framework is learning stable patters in the data. Subnetwork~$1$ highlights regions from the default mode network, which is widely inferred within the resting state literature, and known to play a critical role in consolidating working memory \cite{sestieri2011episodic}. Subnetworks~$1$, $3$ and $4$ highlight regions from the somatomotor network and visual cortex, together believed to be important functional biomarkers of cognitive intelligence \cite{chen2019resting}. Finally, Subnetwork~$2$ and $4$ displays contributions from the frontoparietal network and the medial prefrontal network. These areas are believed to play a role in working memory, attention, and decision making, all of which are associated with cognitive intelligence \cite{menon2011large}.

\medskip\noindent{\textbf{Application to a Secondary Dataset:}} We evaluate our framework on a second clinical cohort of $57$ children with high-functioning Autism Spectrum Disorder (ASD). Rs-fMRI and DTI scans were acquired on a Philips $3T$ Achieva scanner (\textbf{rs-fMRI:} EPI, TR/TE = $2500/30$ms, res = $3.05\times3.15\times3$mm, duration = $128$ or $156$ time samples; \textbf{DTI}: EPI, TR/TE = $6356/75$ms, res = $0.8\times 0.8 \times 2.2$mm, b-value = $700$s/$mm^{2}$, $32$ gradient directions). Rs-fMRI preprocessing includes motion correction, normalization to MNI, spatial and temporal filtering, and nuisance regression with CompCorr \cite{behzadi2007component}. DTI data was preprocessed using the FDT pipeline in FSL \cite{jenkinson2012fsl}, with tractography performed using the BEDPOSTx and PROBTRACKx functions in FSL \cite{behrens2007probabilistic}. We define $\mathbf{y}_n$ using three phenotypes that characterize various impairments associated with ASD: (1) the Autism Diagnostic Observation Schedule (ADOS) \cite{payakachat2012autism}, (2) the Social Responsiveness Scale (SRS) \cite{payakachat2012autism}, and (3) Praxis \cite{mostofsky2006developmental}. We carry forward the same parcellation scheme and model parameters as used for the HCP dataset. Table~\ref{table:KKI} compares the \textit{multi-score} prediction testing performance of ADOS, SRS, and Praxis in a five fold cross validation setting. We observe that only our framework and the model of \cite{d2019integrating} can \textit{simultaneously predict all three measures}. In contrast, the other baselines achieve good testing performance on one or two of the measures (for example, No DTI decoder baseline for ADOS and SRS) but cannot generalize all three. Our supplementary document includes additional results for SC recovery and FC biomarker extraction, and scatter plots for phenotypic prediction on the ASD datatset. Overall, our experiments on both healthy (HCP) and clinical (ASD) populations suggest that our model is robust across cohorts and generalizes effectively even with modest dataset sizes.
\begin{table*}[t!]
\caption{Multi-score performance on the ASD dataset using Median Absolute Error (MAE), Normalized Mutual Information (NMI), and Correlation Coefficient (R) for testing. Best performance is highlighted in bold. Near misses are underlined\label{table:KKI}}
\centering
{\begin{tabular}{|c | c | c | c | c |} 
\hline 
  \textbf{Measure} &\textbf{Method}  & \textbf{MAE Test} & \textbf{NMI Test}  & \textbf{R} \\  
\hline
  \multirow{6}{*}{ADOS} 
 & No rs-fMRI dec. & 3.11~\rpm~{2.74} & \underline{0.46} & 0.089 \\
 & No DTI dec. &  \textbf{2.61~\rpm~{2.59}} & 0.41 & 0.14  \\
 & Decoupled & \textbf{2.64~\rpm~{2.30}} & \textbf{0.49} & 0.35 \\
 & BrainNetCNN & 3.89~\rpm~{2.80} & 0.35 & 0.05 \\
 & Dict. Learn.+ANN & \underline{2.71~\rpm~{2.40}}  & 0.43 & \underline{0.50}  \\
 & \textbf{Our Framework} & \underline{2.71~\rpm~{1.84}} & \textbf{0.49} & \textbf{0.51}  \\
\hline
\multirow{6}{*}{SRS} 
 & No rs-fMRI dec. & 16.84~\rpm~{16.01} & 0.77 & 0.039\\
 & No DTI dec. & \textbf{15.65~\rpm~{12.69}} & 0.81 & 0.24 \\
 & Decoupled & 17.40~\rpm~{14.16} & 0.74 & 0.02 \\
 & BrainNetCNN & 17.50~\rpm~{15.18} & 0.73 & 0.15 \\
 & Dict. Learn.+ANN  & \underline{16.79~\rpm~{13.83}}  & \textbf{0.89} & \textbf{0.37} \\
 & \textbf{Our Framework} & \underline{16.04~\rpm~{13.40}} & \underline{0.83} & \underline{0.34}  \\ 
 \hline
\multirow{6}{*}{Praxis} 
 & No rs-fMRI dec.  & 14.03~\rpm~{10.80} & 0.74  & 0.02 \\
 & No DTI dec. & 19.65~\rpm~{13.18} & 0.81 & 0.23 \\
 & Decoupled  & 17.08~\rpm~{12.23}  & 0.76 & 0.09\\
 & BrainNetCNN  & 19.35~\rpm~{12.56} & 0.74 & 0.20\\
 & Dict. Learn.+ANN  & \textbf{13.19~\rpm~{10.75}} & \underline{0.82} & \textbf{0.37}  \\
 & \textbf{Our Framework} & \textbf{13.14~\rpm~{10.78}} & \textbf{0.86} & \underline{0.32}  \\ 
 \hline
\end{tabular}}
\end{table*}
\section{Conclusion}
We have introduced a novel matrix autoencoder to map the manifold of rs-fMRI functional connectivity to the manifold of DTI structural connectivity. Our framework is strategically designed to leverage the underlying geometry of the data spaces and robustly recover brain biomarkers that are simultaneously explanative of behavioral phenotypes. We demonstrate that our framework offers both interpretability and generalizability, even for multi-score prediction on modest sized datasets. Finally, our framework makes minimal assumptions, and can potentially find application both within and outside the medical realm.
{\paragraph{\textbf{Acknowledgements.}}
This work was supported by the National Science Foundation CRCNS award 1822575, National Science Foundation CAREER award 1845430, National Institute of Mental Health (R01 MH085328-09, R01 MH078160-07, K01 MH109766 and R01 MH106564), National Institute of Neurological Disorders and Stroke (R01NS048527-08), and the Autism Speaks foundation.}

\pagebreak

\bibliographystyle{splncs04}
{{\bibliography{MyRefs.bib}}}
\end{document}